\documentclass[floatfix,groupedaddress,superscriptaddress,aps,showpacs,amsmath,amssymb,floatfix,prl,longbibliography,twocolumn,a4]{revtex4-2}

\usepackage{graphicx,float}
\usepackage{subfigure}
\usepackage{dcolumn}
\usepackage{bm}
\usepackage{mathrsfs}
\usepackage{txfonts}
\usepackage{CJK}
\usepackage[amsmath]{SIunits}
\usepackage{epsfig}
\usepackage{epstopdf}
\usepackage{lipsum}
\usepackage{color}
\usepackage{placeins}

\usepackage[colorlinks,
linkcolor=blue,
anchorcolor=blue,
citecolor=blue,
filecolor=blue,
menucolor=blue,
runcolor=blue,
urlcolor=blue,
frenchlinks=blue]{hyperref}

\allowdisplaybreaks[2]

\allowdisplaybreaks[2]

\begin{document}

\title{A passive bias-free ultrabroadband optical isolator based on unidirectional self-induced transparency}

\author{Haodong Wu}  %
\affiliation{College of Engineering and Applied Sciences, Nanjing University, Nanjing 210023, China}

\author{Jiangshan Tang}  %
\affiliation{College of Engineering and Applied Sciences, Nanjing University, Nanjing 210023, China}

\author{Mingyuan Chen}  %
\affiliation{College of Engineering and Applied Sciences, Nanjing University, Nanjing 210023, China}


\author{Min Xiao}
\affiliation{School of Physics, Nanjing University, Nanjing 210023, China}
\affiliation{Department of Physics, University of Arkansas, Fayetteville, Arkansas 72701, USA}

\author{Franco Nori}
\email{fnori@riken.jp}
\affiliation{RIKEN Quantum Computing Center, RIKEN Cluster for Pioneering Research, Wako-shi, Saitama 351-0198, Japan}
\affiliation{Physics Department, The University of Michigan, Ann Arbor, Michigan 48109-1040, USA}

\author{Keyu Xia }
\email{keyu.xia@nju.edu.cn}
\affiliation{College of Engineering and Applied Sciences, Nanjing University, Nanjing 210023, China}
\affiliation{National Laboratory of Solid State Microstructures, Nanjing University, Nanjing 210023, China}
\affiliation{Hefei National Laboratory, Hefei 230088, China}
\affiliation{Shishan Laboratory, Suzhou Campus of Nanjing University, Suzhou 215000, China}

\author{Yanqing Lu}
\email{yqlu@nju.edu.cn}
\affiliation{College of Engineering and Applied Sciences, Nanjing University, Nanjing 210023, China}
\affiliation{National Laboratory of Solid State Microstructures, Nanjing University, Nanjing 210023, China}

\date{\today}

\begin{abstract}
Achieving a broadband nonreciprocal device without gain and any external bias is very challenging and highly desirable for modern photonic technologies and quantum networks. 
Here, we theoretically propose a passive and bias-free all-optical isolator for a femtosecond laser pulse by exploiting a new mechanism of unidirectional self-induced transparency, obtained with a nonlinear medium followed by a normal absorbing medium at one side. The transmission contrast between the forward and backward directions can reach $~14.3~\deci\bel$ for a $2\pi$ $5~\femto\second$ laser pulse, implying isolation of a signal with a ultrabroad bandwidth of $200~\tera\hertz$. The $20~\deci\bel$ bandwidth is about $57~\nano\meter$, already comparable with a magneto-optical isolator. This cavity-free optical isolator may pave the way to integrated nonmagnetic isolation of ultrashort laser pulses.
\end{abstract}

\maketitle

\section{Introduction}Nonreciprocal optical devices (NRODs), such as optical isolators and circulators, enforcing unidirectional light propagation, have become key components for modern photonic applications~\cite{Millot2016NP, Zhang2022Nture, Jiho2019Science, Lai2020NP}, quantum information processing~\cite{Daiss2021, Wallucks2020, 2008The, Cirac1997, Andre2017}, quantum sensing~\cite{PhysRevA.103.042418} and nanofabrication of optical crystals~\cite{ZhangYongNature}.
NRODs need to break the electromagnetic time-reversal symmetry. A standard approach to implement NRODs is based on the Faraday effect using a DC magnetic bias in a magneto-optical material~\cite{Ren2022}. However, a magneto-optical system is typically incompatible with integrated photonic technologies and magnetically sensitive applications, introduces large loss to the signal, and requires a strong magnetic field. 

Various schemes for magnetic-free nonreciprocity have been proposed and demonstrated by using spatiotemporal modulation~\cite{Lira2012,Estep2014,Sounas2017, NaturePhotonics.8.701, NaturePhotonics.15.828}, nonlinear resonator modes~\cite{RN43, Fan2012, Alu2022arxiv}, quantum nonlinearity~\cite{ZTC2019PRL, AtomicMirrorPRL, Blatt2011PRL}, nonlinearity in a parity-time-symmetry-broken system~\cite{JXS2014NP, YL2014NP, XM2018PRL, PhysRevA.99.053806}, photon-phonon coupling~\cite{Kittlaus2021, DCH2016NP, 2011NP, Gaurav2018NP, Gaurav2015NP}, moving lattice~\cite{WDW2013PRL, WJH2013PRL}, spinning resonators~\cite{JH2018Nature, JH2020PRL, JH2018PRL}, wave-mixing processes in nonlinear media~\cite{FSH2009NP, Song2021, Yang2022, JXS2016NC}, chiral quantum optical systems~\cite{XKY2014PRA, Arno2016PRX, Arno2016Science, Peter2015NNT, TL2019PRA,  Tang2022PRL, XKY2018NP, XKY2018PRL, XKY2021SciAdv, XKY2020PRL,  WHD2022LPR, ZCL2021NC, arxiv2210.07038, Philipp2022NP, TL2022AQT}, chiral valley systems~\cite{Guddala2021} and unidirectional quantum squeezing of microring resonator modes~\cite{Tang2022a, ZCL2016PRL}. In addition, phononic and microwave nonreciprocity have been reported and have their own applications~\cite{Xu2020,Li2011, Wang2019, Huang2016,PhysRevApplied.10.064037}. Meanwhile, these approaches require high-quality cavities to enhance the light-matter interaction or narrow-band auxiliary systems, such as alkali atoms or mechanical modes. As a result, their nonreciprocal bandwidth is strongly limited.

Nonlinearity-based NRODs have attracted significant attention because they are compatible with integrated photonics and can be obtained in the absence of an external bias. Nevertheless, dynamic reciprocity imposes fundamental constraint to their practical applications~\cite{Shi2015}. The chiral Kerr nonlinearity can tackle the problem of dynamic reciprocity~\cite{XKY2018PRL, SBS2020PRR, Leonardo2018Optica, PanRuikaiCPL}. However, nonlinear NRODs often require nonlinear resonators, which have a very limited bandwidth.

In this theoretical work, we propose a passive bias-free optical isolator for a ultrashort laser pulse with a nonlinear medium with asymmetric inputs exhibiting unidirectional self-induced transparency (SIT). This cavity-free isolator can enable unidirectional propagation of a $5~\femto\second$ laser pulse. A $57~\nano\meter$ bandwidth can be obtained for $20~ \deci\bel$ isolation.

\section{System and model}
\begin{figure}
  \centering
  \includegraphics[width=1.0\linewidth]{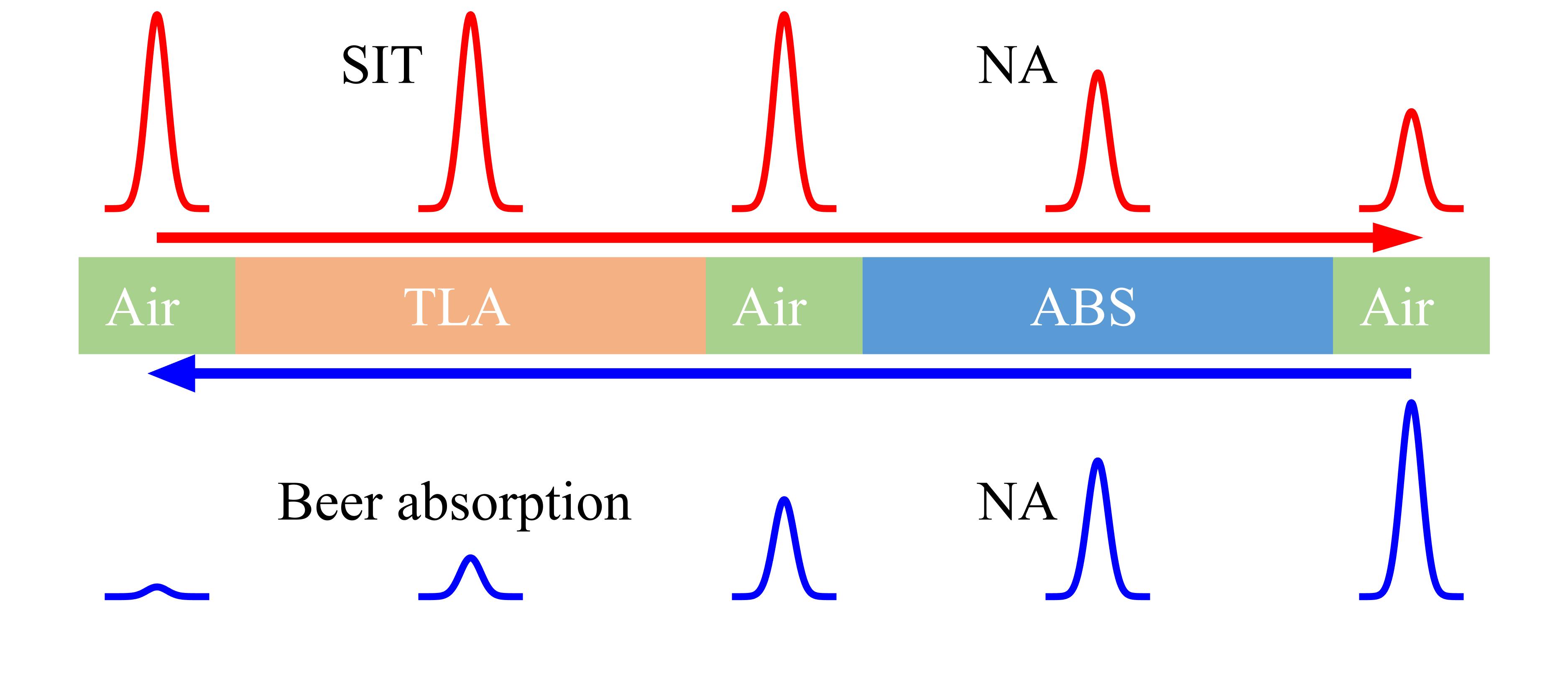} \\
\caption{Schematic of the proposed optical isolator consisting of a TLA medium and a NA medium. The forward-moving femtosecond laser pulse first passes the TLA medium without loss due to the SIT and then is partly absorbed by the NA medium (red pulses). The backward-moving pulse (blue pulses) first decays to an area smaller than $\pi$ due to the absorption in the NA medium, and is further strongly absorbed in the TLA medium according to Beer's law~\cite{AllenBook}.}
\label{fig1}
\end{figure}

Our proposed optical isolator consists of a 1D two-level atom (TLA) medium and a normal-absorpting (NA) medium, as schematically depicted in Fig.~\ref{fig1}. The TLA medium is followed by an air region and then by a NA medium in the right-hand end. The NA medium causes absorption to a laser field. For simplicity, we consider the TLA and NA media to be surrounded by air, for the theoretical analysis. In practice, the TLA medium can be a semiconductor~\cite{Polu1975Self,Wu2011}, an ensemble of atoms~\cite{Gibbs1972}, or semiconductor quantum dots doped in an optical material~\cite{PRL94,PRL81,APL10,Kim2022}. The NA medium can immediately follow the TLA medium. The air can also be replaced with a transparent optical medium.

We use a two-level model for the TLA medium. The transition between the ground state $|g\rangle$ and the excited state $|e\rangle$ of the TLA system has a resonance frequency $\omega_0$ and a electronic dipole moment $\vec{d}$ with amplitude $d$. The population relaxation time and the decoherence time are $T_1$ and $T_2$, respectively.

We assume that a laser field is polarized along $\vec{d}$ with angular frequency $\omega_L$. At position $z$ and time $t$, its envelope is $\tilde{E}_x(z, t)$, with its area defined as $A = \int_{-\infty}^\infty d \tilde{E}_x(t)/\hbar dt$~\cite{AllenBook}. The pulse energy is calculated as $U_p(z) = \frac{ \varepsilon_0}{2} \int_{-\infty}^\infty \tilde{E}^2_x(z, t) dt$, with $\varepsilon_0$ being the vacuum permittivity.

The propagation of a short laser pulse with a duration $\tau_p \ll T_2$ in a TLA medium is crucially dependent on its area $A$~\cite{AllenBook}. Due to the strong nonlinear interaction, an ultrashort laser pulse with $A = 2\pi$ and $\omega_L = \omega_0$ drives a full Rabi oscillation of the TLA medium. This process takes place much faster than the relaxation and decoherence rates. During propagation, the first-half part of the $2\pi$ laser pulse coherently transfers energy to atoms and completely inverts the atomic population. Then, the atoms coherently return back the energy to the laser pulse after being ``pulled'' back to the ground state by the second-half part. Thus, the $2\pi$ laser pulse can propagate as the nonlinear medium is transparent and maintain its area $2\pi$. Thus, the laser pulse can be considered as a light soliton with very small energy loss. This SIT phenomenon has been demonstrated in various experiments~\cite{Carmon2000OL, DN2003OL, Rotschild2006OL} and by numerical simulations based on the Maxwell-Bloch equations~\cite{McCall1967PRL, Hughes1998PRL, Daniel1995PRA, Xia2005, Xia2007, Xia2007a, xia2007b, xie2010}.  When $\pi < A < 2\pi$,  the pulse will evolve to a $2\pi$ pulse at the expense of losing some energy. If $A < \pi$, the pulse will be strongly absorbed. Its total intensity or energy reduces exponentially according to Beer's law~\cite{AllenBook}. A $\pi$ pulse is unstable and will decay during propagation.

The key idea of our isolator is explained as follows. When an ultrashort pulse with $A \sim 2\pi$ enters the device from the left end in the forward case, the TLA medium is mostly transparent because of the SIT. Then, the pulse leaves the device after a small absorption in the NA medium. This forward transmission can be high. In contrast, in the backward case, the pulse first inputs to the right end of the NA medium. It is partly absorbed by the NA medium and then suffers a Beer-law decay in intensity (energy) inside the TLA medium. Thus, the asymmetric arrangement of the NA medium causes unidirectional SIT, indicating a strong bias-free optical nonreciprocity.

Now we model the propagation of the laser pulse $E_x (t)$ in the TLA medium with the Maxwell-Bloch coupling equations. The TLA-field interaction  can be described by the Hamiltonian without the rotating-wave approximation and the slowly varying envelope approximation as
\begin{equation}
H = \left(
\begin{matrix}
\omega_0/2 & d E_x(t)/\hbar \\
d E_x(t) /\hbar   & - \omega_0/2
\end{matrix}
 \right)\;.
\end{equation}
We consider a 1D continuous TLA medium with a number density $N_a$. The Maxwell's equations for the electric field $E_{x}$ and the magnetic field $B_y$ take the form
\begin{subequations}\label{eq:Maxwell}	
\begin{align}
     \partial_t B_y & = - \partial_z E_x \;,\\
     \partial_t D_x & = -\frac{1}{\mu_0}\partial_z B_y \;,
\end{align}
\end{subequations}
where $\mu_0$ is the magnetic permeability in the vacuum. The nonlinear response of the TLA medium is taken into account via the relation $D_x = \varepsilon_0 E_x + P_x$, where the macroscopic polarization $P_x = - N_a d u$ is connected with the off-diagonal density matrix element $\langle e| \rho |g\rangle = (u + i v)/2$. $P_x$ can be solved by the Bloch equations derived from $H$
\begin{subequations}\label{eq:Bloch}	
\begin{align}
	\partial_t u & = -\frac{1}{T_2}u + \omega_0 v \;,\\
	\partial_t v & = - \omega_0 u -\frac{1}{T_2}v + 2\frac{d E_x}{\hbar} w \;,\\
	\partial_t w & = - 2\frac{d E_x}{\hbar} w - \frac{1}{T_1}(w-w_0) \;,
\end{align}
\end{subequations}
where $w$ is the population difference between the states $|g\rangle$ and $|e\rangle$. Note that $u$, $v$, and $w$ satisfy the relationship $u^2+v^2+w^2=1$.
This model can reproduce well experimental results~\cite{AllenBook, rfSIT} and has been widely used to study the time evolution of a femtosecond laser pulse in a nonlinear medium~\cite{RN2, FX2002PRL}.

\section{Numerical method}We employ the standard finite-difference time-domain method (FDTD) and the fourth-order Runge-Kutta method to solve the Maxwell-Bloch coupling equations~\cite{Hughes1998PRL, Xia2007, Xia2007a, xia2007b}. The Mur absorbing boundary condition is used to avoid the reflection off the truncated boundary of the FDTD simulation domain~\cite{Mur1981}. 
A source laser pulse is introduced to the air region at $z_\text{f} = 15~\micro\meter$ in the forward case and at $z_\text{b} = 255~\micro\meter$ in the backward case. It is assumed to be a sech function defined as $E_x(t) = \tilde{E}_x(t,\tau) \sin (\omega_\text{L} (t-\tau))$ with $\tilde{E}_x(t, \tau) = E_0 \text{sech}[1.76 (t-\tau)/\tau_p]$, where $E_0$ is the amplitude. The delay $\tau$ is much larger than duration $\tau_p$ to ensure that the electric field is negligible at $t=0$. Specifically, $\tau > 4 \tau_p$. The area of the input pulse is $A_0 = dE_0 \tau_p \pi/1.76\hbar$.

Now we specify the TLA medium and the NA medium. The TLA medium is set at $30\leq z \leq 120~\micro\metre$ and the NA medium at $150 \leq z \leq 240~\micro\metre$. These are separated by $30 ~\micro\meter$ of air, so that we can monitor the transmitted field after the media. The medium in the rest regions is air. The time-dependent transmitted pulses are retrieved at $z = 15~\micro\meter$ in the backward case and at $z = 255~\micro\meter$ in the forward case. The NA medium causes a decay of $2.95~\deci\bel$ in the total field energy.
We choose the atomic parameters: $N_a = 3\times10^{25}~{\metre}^{-3}$ $\omega_0=2.3\times10^{15}~\hertz$  ($\lambda_0=820~\nano\meter$), $d=2\times10^{-29}~\coulomb \cdot \meter$, $T_1^{-1}=0.5~\pico\second$ and $T_2^{-1}=0.25~\pico\second$. The TLA medium is initialized in the ground state such that $u=0$, $v=0$, and $w_0= -1$ at $t=0$.

\begin{figure}
	\centering
	\includegraphics[width=1.0\linewidth]{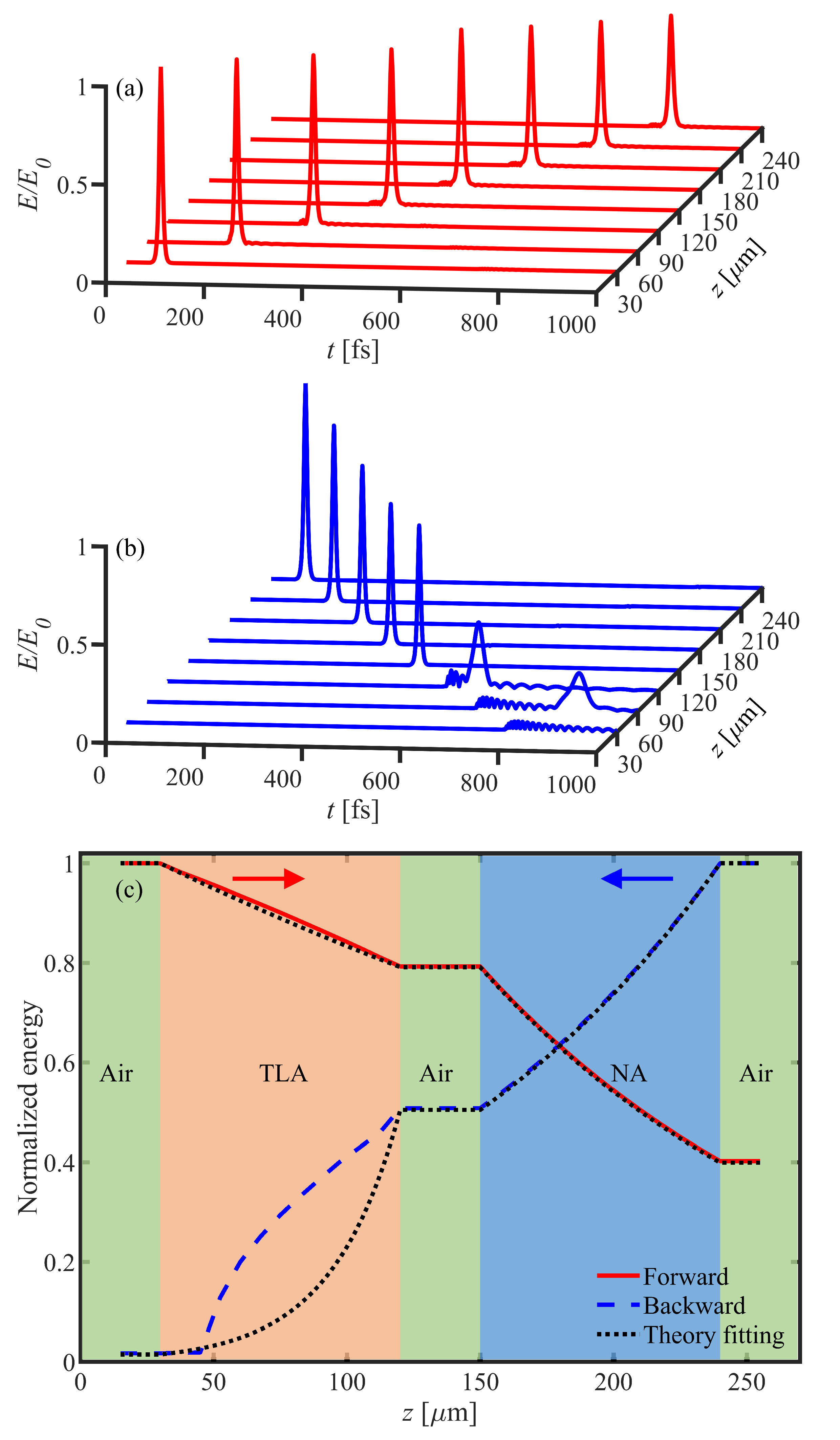} \\
	\caption{Evolution of a $2\pi$ femtosecond laser pulse with $\tau_p = 5~\femto\second$ in the forward case (a) and the backward case (b). (c) Energy evolution of the forward- (red curve) and backward-moving (blue cuve) laser pulses. Dotted curves are fits with exponential decays.}
	\label{fig2}
\end{figure}
Without loss of generality, we adopt the typical duration $\tau_p = 5~\femto\second$ for a resonant femtosecond laser pulse ($\omega_\text{L} = \omega_0$), spanning a $200~\tera\hertz$ bandwidth. For $\tau_p = 5~\femto\second$, the pulse area can reach $A_0=2\pi$ when $E_0 \approx 3.7\times10^9~\volt\per\meter$. We are interested in a laser pulse with $A_0 < 3\pi$, corresponding to a peak field intensity less than $4\times 10^{12}~\watt/{\centi\meter}^{2}$. 

We evaluate the performance of the optical isolator by comparing energies of the input and transmitted laser fields. The forward and backward transmissions are calculated as 
\begin{subequations} \label{eq:Transmissions}
\begin{align}
T_\text{f} = -10 ~{\log}_{10} [U(z=240~\micro\metre)/U(z=30~\micro\metre)] \;, \\
T_\text{b} = -10 ~{\log}_{10} [U(z=30~\micro\metre)/U(z=240~\micro\metre)] \;.
\end{align}
\end{subequations}
In calculation of the pulse energy and the transmissions, we truncate the integral time limits because the energy of the laser pulses outside the limits is negligible. The isolation contrast is defined as $\eta = 10 {\log}_{10} (T_\text{f}/T_\text{b})$.

 \begin{figure}
	\centering
	\includegraphics[width=1\linewidth]{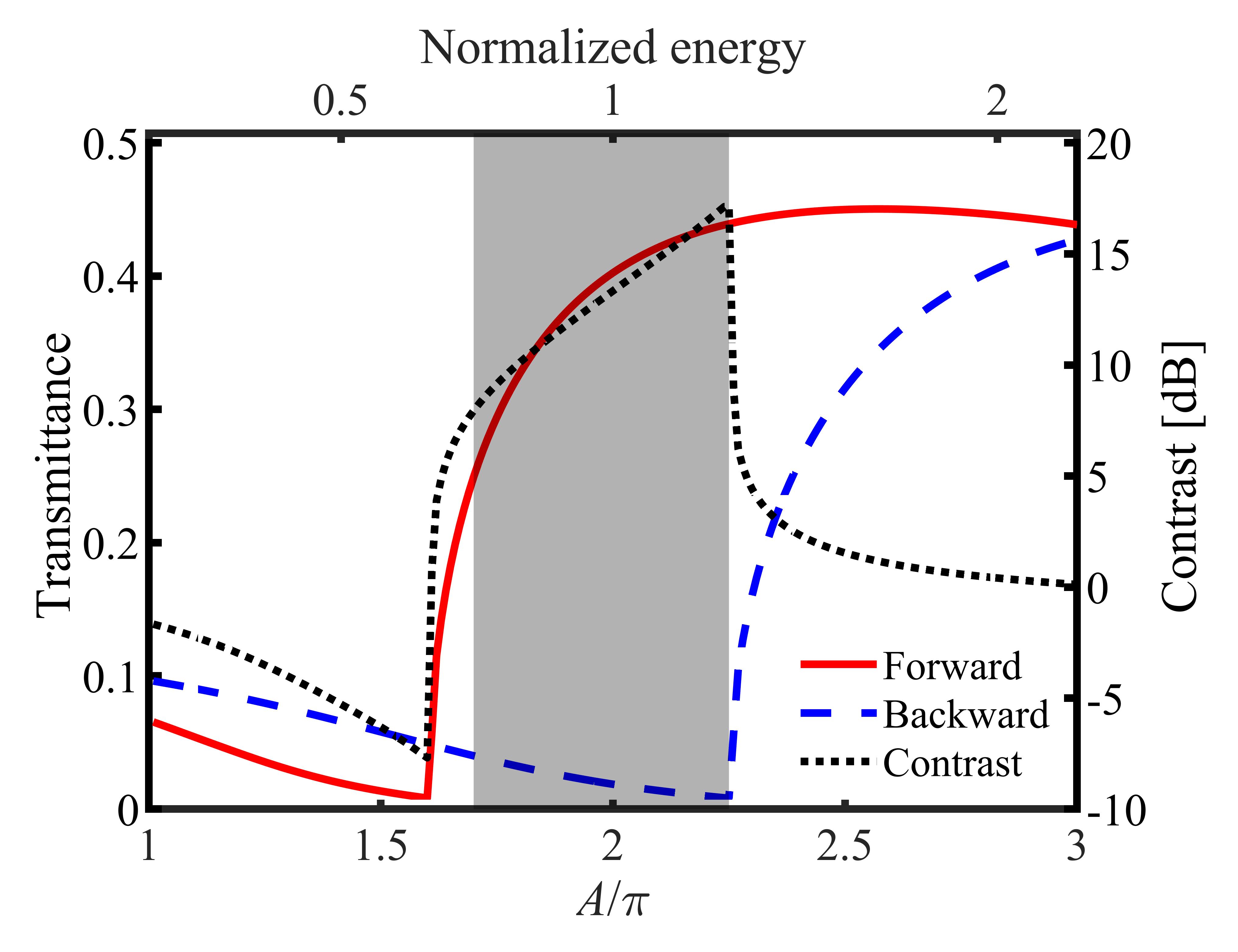} \\
	\caption{ Transmission and isolation contrast of the optical isolator versus the pulse area A and energy for a $5~\femto\second$ laser pulse.}
	\label{fig3}
\end{figure}
\section{Theoretical Results}We first investigate the propagation of a $2\pi$ laser pulse excited by the source at either $z_\text{f}$ or $z_\text{b}$, respectively. The nonreciprocal propagation can be seen from Fig.~\ref{fig2}. In the forward case, see Fig.~\ref{fig2}(a), the laser pulse enters the TLA medium at about $t=100~\femto\second$ with a negligible reflection and leaves at $t=400~\femto\second$ with the unchanged pulse shape. The laser energy decays to $79.1\%$ of the initial energy $U_0$ due to the fast atomic dissipation. After going through the nonlinear medium, the energy of the pulse further decays by about $2.95~\deci\bel$ due to the absorption of the NA medium. The laser pulses transmitted through the device retain $40.2\%$ of their energy, i.e. $T_\text{f} = 40.2\%$. Nevertheless, the pulse keeps its shape unchanged after passing through the isolator. In the backward case, see Fig.~\ref{fig2}(b), the laser pulse with the same area is excited by the source at $z_\text{b}$. In stark contrast, the backward-moving laser pulse is first absorbed to an area well smaller than $2\pi$, corresponding to $50.8\%$ of the initial energy, and then enters the TLA medium at $t = 600~\femto\second$. The pulse area remains about $1.4\pi$. During propagation, the pulse quickly splits into two parts, a fast-decaying strong pulse and a slowly-decaying weak one with an oscillating envelope. According to Beer's law, the TLA medium strongly absorbs the main pulse but leaves the negligibly small oscillating one, corresponding to a $2\pi$ pulse but with a smaller energy decay scale with respect to the Beer law. When the laser field passes through the TLA medium at $t = 900~\femto\second$, only $1.48\%$ of the energy of the pulse remains, corresponding to $T_\text{b} = 1.48\%$. 
This nonreciprocal energy evolution of the laser pulses due to the unidirectional SIT during propagation is clearly shown in Fig.~\ref{fig2}(c). The obtained isolation contrast and the insertion loss are $14.3~\deci\bel$ and $3.96~\deci\bel$, respectively. The insertion loss in the forward case dominantly results from the absorption in the NA medium.

\begin{figure}
\centering
\includegraphics[width=1.0\linewidth]{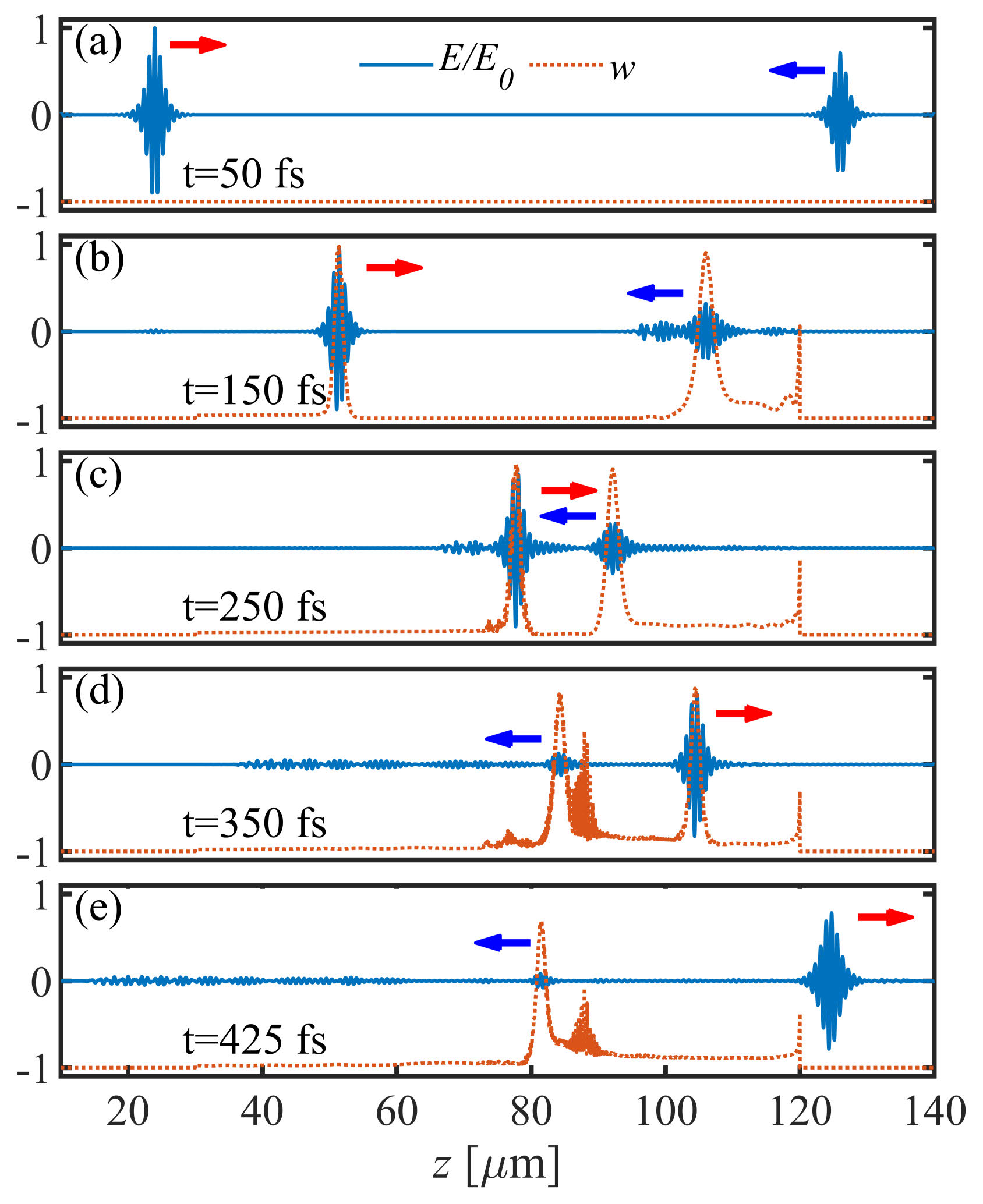} \\
\caption{Field distribution of a $5~\femto\second$ laser pulse (blue curves) in the TLA medium and the population difference $w$ (brown curves) at different times. Red and blue arrows indicate the forward and backward moving directions.}
\label{fig4}
\end{figure}
Therefore, our theoretically proposed optical isolator can enable nonreciprocal propagation of a femtosecond laser pulse spanning over $200~\tera\hertz$. The discrepancy between the energy evolution (simulations) and an exponential decay (fitting) in the TLA medium is due to the slow-decaying and envelope-oscillating $2\pi$ pulse, which includes a small fraction of the total energy.

The transmission and isolation contrast versus the area of the input laser pulse are shown in Fig.~\ref{fig3}. Here, the isolation contrast is calculated for a whole femtosecond laser pulse.
When $2\pi < A < 3\pi$, the forward transmission is stable and high, larger than $40\%$. We obtain the maximal isolation contrast $17.1~\deci\bel$ when $A = 2.25 \pi$. However, when $A>2.25 \pi$, the backward transmission quickly increases to a high level. According to the ``area theorem''~\cite{AllenBook}, if the area of a femtosecond laser pulse is less than $\pi$ when it enters the TLA medium, its energy can be completely absorbed in the TLA medium. However, when $A>2\pi$, the laser pulse will split into multi daughter pulses although the daughter pulses are under the SIT. To guarantee a high and shape-maintaining transmission in the forward case, the  pulse area when entering the TLA medium needs to be no more than $2\pi$ to avoid splitting. Thus, the nonreciprocal energy range is narrow. Note that there is a small reflection caused by the air-medium interface. We can attain high-performance isolation for $1.81 \pi < A< 2.25 \pi$.

Dynamic nonreciprocity is also obtained when the two laser pulses simultaneously propagate in the TLA medium along opposite directions. We study the two counter-propagating $2\pi$ laser pulses as an example. To allow the two laser pulses to meet in the medium, we input the backward-moving pulse about $500~\femto\second$ earlier than the forward-moving pulse. Figure~\ref{fig4} shows the normalized electric fields of the pulses and the population inversion $w$. When the forward- and backward-moving pulses meet, the population inversion $w$ oscillates quickly. However, the two laser pulses can pass each other without observable interference as two light solitons. The transmission of the backward-moving pulse is much smaller than the forward-moving pulse. These results confirm that our proposed \emph{optical isolator displays dynamic nonreciprocity}.
\begin{figure}
	\centering
	\includegraphics[width=1\linewidth]{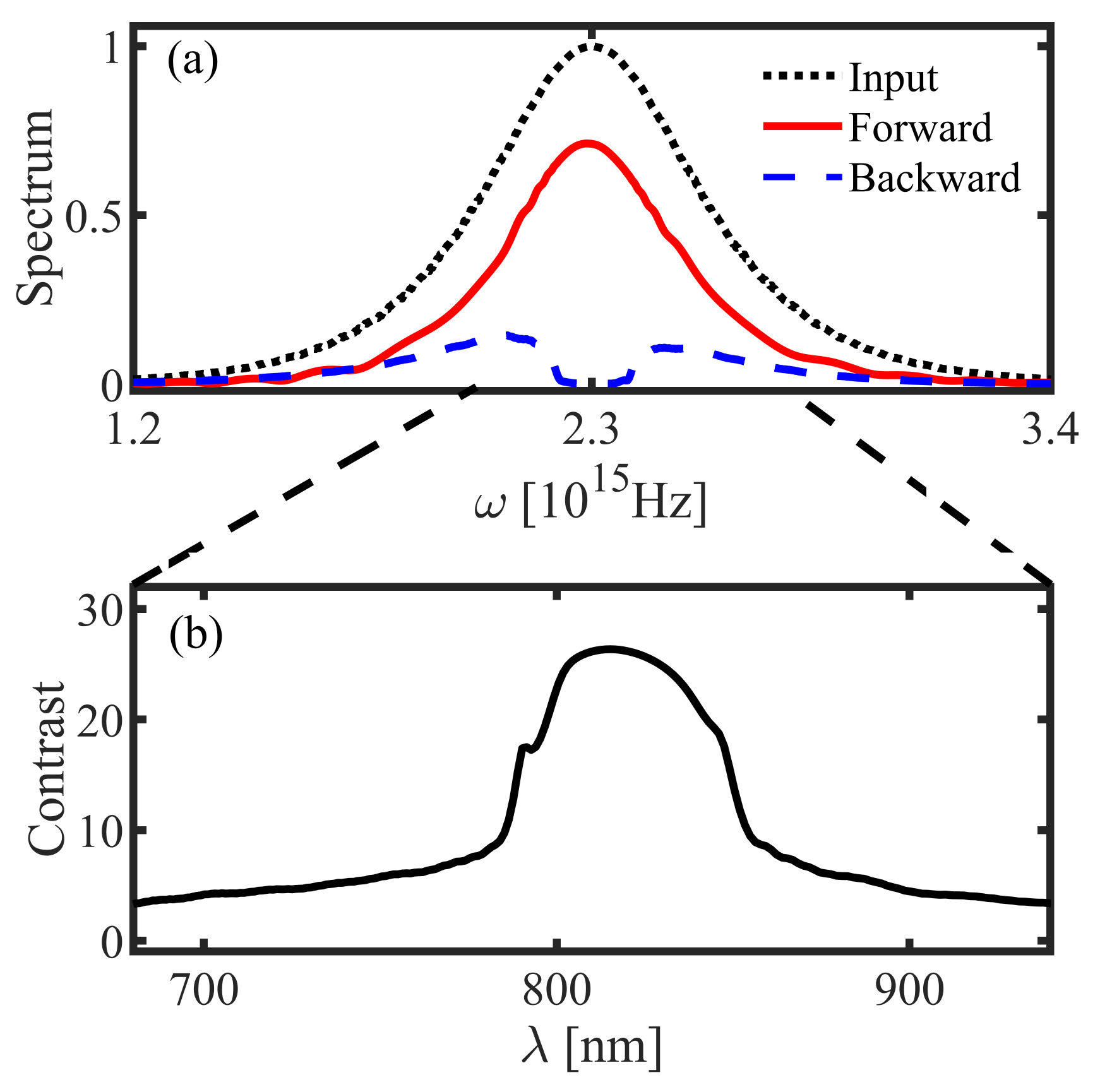} \\
	\caption{(a) Spectra of the input pulse (dotted black curve), the forward-moving (solid red curve) and the backward-moving (dashed blue curve) pulses transmitted through the optical isolator. (b) Isolation contrast versus the wavelength.}
	\label{fig5}
\end{figure}

We study the transmission spectra and the isolation contrast versus the frequency (wavelength) in Fig.~\ref{fig5}.
We can obtain the pulse spectra by the Fourier transform of the corresponding fields. The spectra are normalized with respect to the peak value of the input pulse spectrum. We take the time when the laser field disappears as the lower and upper temporal boundaries of the Fourier transform. As shown in Fig.~\ref{fig5}(a), after transmitting the optical isolator, the forward-moving laser pulse retains a spectral profile similar to the input pulse, whereas a deep dip appears in the spectral center of the backward-moving pulse due to the strong Beer-law absorption of the TLA medium. The isolation contrast around the dip is shown in Fig.~\ref{fig5}(b). We obtain a maximum contrast $28.1~\deci\bel$ at $\omega_\text{L}$ and a nonreciprocal bandwidth of $57~\nano\meter$ for $\eta \geq 20~\deci\bel$. The isolation can be improved to a much higher value by using a TLA medium long enough, because the decay of the forward-moving laser pulse is mostly determined by the NA medium and thus fixed, while the backward-moving laser pulse with an area $A<\pi$ initially in the TLA medium can be completely absorbed.
Here, we choose a $90~\micro\meter$ nonlinear medium for demonstrating the concept of our isolator and remaining the reasonably high forward transmission.

\section{Implementation}The SIT has been observed in various media, such as Ruby~\cite{Asher1972}, alkali atoms~\cite{Gibbs1972}, Rydberg atoms~\cite{LWB2020PRL}, molecular gases~\cite{Zembrod1971}, semiconductors~\cite{Polu1975Self,Wu2011} and semiconductor quantum dots~\cite{PRL94, PRL81, PRB65, APL10}. Semiconductors possess short relaxation times and comparatively high transition dipole moments. Semiconductor quantum dots can be artificial TLAs with high dipole moments~\cite{PRB65, APL10}. The conditions required by our scheme can be satisfied with semiconductors~\cite{PRL94, PRL81}.
Thus, an on-chip optical isolator based unidirectional SIT can be integrated on a semiconductor platform.

\section{Conclusion and discussion}In summary, we have shown unidirectional SIT as a novel mechanism for obtaining a passive bias-free optical isolator. This isolator displays dynamic nonreciprocity over an ultrabroad bandwith comparable with a magneto-optical nonreciprocal device. This ultrabroadband nonreciprocial device has the potential to be integrated on a chip in a semiconductor platform and thus can boost applications of integrated photonics and femtosecond lasers. 

Note that SIT of a few-cycle rf pulse has been experimentally observed in Rubidium atoms and is reproduced by solving the Bloch equation~\cite{rfSIT}. Therefore, the concept of  our optical isolator  can be extended to the rf regime. If a TLA medium with small dissipation is available~\cite{LWB2020PRL}, our method can also isolate the reflection of a long light pulse.

\section*{Acknowledgements}
H.W. thanks Lei Tang for helpful discussions.
This work was supported by the National Key R\&D Program of China (Grants No. 2019YFA0308700 and No. 2019YFA0308704), the National Natural Science Foundation of China (Grant No. 11874212 and No.11890704), the Fundamental Research Funds for the Central Universities (Grant No. 021314380095), the Innovation Program for Quantum Science and Technology (Grant No. 2021ZD0301400), the Program for Innovative Talents and Teams in Jiangsu (Grant No. JSSCTD202138).
F.N. is supported in part by: Nippon Telegraph and Telephone Corporation (NTT) Research, the Japan Science and Technology Agency (JST) [via the Quantum Leap Flagship Program (Q-LEAP), and the Moonshot R\&D Grant Number JPMJMS2061], the Japan Society for the Promotion of Science (JSPS) [via the Grants-in-Aid for Scientific Research (KAKENHI) Grant No. JP20H00134], the Army Research Office (ARO) (Grant No. W911NF-18-1-0358), the Asian Office of Aerospace Research and Development (AOARD) (via Grant No. FA2386-20-1-4069), and the Foundational Questions Institute Fund (FQXi) via Grant No. FQXi-IAF19-06.
%
%


%

\end{document}